# Universal properties of modulated antiferromagnetic systems

C. D. Batista, G. Ortiz, & A. V. Balatsky

*Theoretical Division, Los Alamos National Laboratory, Los Alamos NM 87545, USA*

**Magnetism and superconductivity are physical phenomena whose foundations are rooted in quantum mechanics and whose technological applications do not cease to surprise us. In this article we describe a class of magnetic materials, we call *modulated antiferromagnetic systems* (MAS), that has a prominent representative in the copper-oxide high-temperature superconductors. The class, however, is not exclusive to these superconductors. Indeed, several materials that belong to that class are insulators. The magnetic spectral weight of MAS displays the following universal properties: a local intensity maximum (*resonance* peak) at the commensurate antiferromagnetic wave vector; peaks at the nearly-antiferromagnetic wave vectors for excitation energies well below the resonance, and at wave vectors rotated by 45 degrees for energies above resonance. Moreover, we predict an observable rotation by 45 degrees of the peaks *immediately below* the resonance. All these universal signatures, condensed in a *twisted hour-glass-like* spectrum, are merely consequences of the unique topological characteristics of the single-particle magnetic dispersion relation. We thus provide a unifying scenario that explains the phenomenology that has been observed in inelastic neutron scattering experiments of the high-temperature superconductors.**

The exchange of spin fluctuations is thought to be one of the possible mechanisms leading to high critical superconducting temperatures (high-$T_c$) in the cuprates. With $T_c$ ~150 Kelvin these materials have the highest transition temperatures known to date. This is the reason why the nature of pairing and the role of spin fluctuations are at the centre of research. The logic is that if we know the spectrum of the intermediary bosonic particles, i.e., spin fluctuations, we might better understand the pairing and the reason for such a robust superconducting state. On the other hand, inelastic neutron scattering experiments (INS) have (almost) systematically revealed the two most striking features in the spin fluctuation spectrum: (doping-dependent) *incommensurate* magnetic excitations[1-5], and a so-called *resonance* peak[6-9] considered a central hallmark that could hypothetically be related with the pairing *glue*. More recently, a new characteristic has been identified, namely the maximum intensity peaks in the magnetic spectral weight seen below the resonance energy *rotate* by 45° as they move to higher excitation energies.

Five years ago[10,11], we argued that the *resonance* peak observed in several high-$T_c$ superconductors has a magnetic origin and that it is a natural and universal consequence of the low-energy incommensurate magnetic fluctuations that appear close to the $Q=(\pi,\pi)$ antiferromagnetic wave vector. Moreover, we stated that any system with a magnetic dispersion relation exhibiting minima at wave vectors $\tilde{Q}_\pm = Q \pm \frac{\pi}{a}\delta \hat{k}$, where $\hat{k}$ is a unit vector and $\delta \ll 1$ (MAS), should exhibit a peak in the magnetic structure factor $S(k,\omega)$ at $k=Q$, and $\omega = E_r$. The peak results from the superposition of the two low-energy magnetic branches that emerge from $\tilde{Q}_\pm$ and converge at a saddle point at $k=Q$ (see Fig. 1). The actual value of the energy $E_r$ depends on the slope of the low-energy branches (spin velocity). Based on this observation, we predicted that a similar peak should be observed in insulating systems that exhibit modulated antiferromagnetic order like $La_{1.69}Sr_{0.31}NiO_4$. This prediction was later confirmed by Bourges *et al.*[12]. Moreover, successive measurements of different high-$T_c$ compounds[13-26] are confirming our idea that a universal magnetic spectrum plus a spin gap can explain various observations in the cuprates [10,11,27].

Despite our arguments being repeatedly confirmed by INS, last findings were interpreted as evidence against our predictions[23]. Two independent groups[22,23] observed that the highest intensity high-energy magnetic excitations (above the resonance energy) are rotated 45° from the incommensurate wave vectors. This behaviour was observed in two different compounds $La_{1.875}Ba_{0.125}CuO_4$[23] and $YBa_2Cu_3O_{6.6}$[22]. While the first compound exhibits long-range charge and spin ordering[28], the latter only exhibits short range dynamical correlations. In this article we will show that our theory not only contains the observed *twist* in the high-energy excitations but also predicts many other insulating systems depicting the same *saddle-point* effect. We will show that this saddle point is another universal feature of an incommensurate spin state and not a unique feature of high-temperature superconductors, as has been frequently emphasized. Moreover, since this feature is universal, its observation is not sufficient to discriminate between competing theories that include incommensurate magnetic correlations (such as weakly-coupled 2-leg ladder[23], etc.). Besides explaining the twist, we predict additional universal behaviours that should be observable both in the cuprates as well as in the magnetic insulators that exhibit incommensurate magnetic ordering.

The universal behaviours of MAS are a simple consequence of a few qualitative aspects of the magnetic dispersion relation. This is illustrated by Fig. 1 which shows the qualitative aspects that emerge from the intersection of four cones centred on each incommensurate wave vector (the four low-energy modes result from averaging over the horizontal and vertical orientations, mimicking a twinned crystal). It is important to note that an isotropic conical shape of the low-energy dispersion is *only expected* for the insulating materials. As it is explained below, this is no longer true for metals in which the magnetic branches that move away from the commensurate point $Q$ can be over-damped by the particle-hole continuum of excitations. This effect, as well as the presence of a spin gap or a possible anisotropy of the conical shape *do not* change the following qualitative properties:    (a) The four cones converge at a saddle point at

$k=Q$ and the corresponding increase in the spectral weight leads to the observed *resonance* peak[10,11]; (b) For higher energies, the intersections between pairs of adjacent cones leads to four points with higher intensity that are *rotated* 45° relative to the incommensurate wave vectors; (c) A similar *twist* should be observed for energies $0.7 E_r \lesssim \hbar\omega \lesssim 0.9 E_r$. While (a) and (b) are experimental facts, (c) is a *new* prediction that requires experimental confirmation.

Since these qualitative properties are universal to any MAS, it is enough to compute $S(k,\omega)$ for a particular system in order to see how the mechanism illustrated in Fig. 1 develops in the INS spectrum. The simplest case corresponds to a magnetic insulator (only spin degrees of freedom). As we did in our previous papers[10,11], the corresponding magnetic spectrum can be obtained using a simple Schwinger-boson mean-field description. The result of this calculation is shown in Fig. 2 (the spin gap is $\Delta_s = 0.1 E_r$). At energies $\omega \ll E_r$, the intensity is centred around the four incommensurate points. For $\hbar\omega \gtrsim 0.7 E_r$, the intersection between adjacent circles leads to regions of higher intensity along the diagonal directions. The four intersection points that are closer to the centre move towards $Q$ with increasing energy until they merge at $\hbar\omega = E_r$, giving rise to the *resonance* peak. For $\hbar\omega > E_r$, the same points continue moving away from $Q$ and leading to the observed *twist* in the high-energy magnetic excitations. Assuming that a linear dispersion relation is still valid in the high-energy region, we can compute the distance $\Delta k$ between these points and $Q$ as illustrated in the inset of Fig. 3. This calculation gives $\Delta k = \frac{\pi}{a}\delta\left(\frac{\sqrt{7}-1}{\sqrt{2}}\right) \sim 1.164 \frac{\pi}{a}\delta$ for $\hbar\omega = 2 E_r$ and the resulting value can be considered as a lower bound for systems in which the curvature in the dispersion relation becomes appreciable at that energy. In particular, for the Schwinger-boson calculation we obtain $\Delta k = 1.4 \frac{\pi}{a}\delta$ in good agreement with the neutron scattering measurements in $La_{1.875}Ba_{0.125}CuO_4$[23].

Finally, it is also important to analyse the non-universal aspects of MAS. As stated above, these systems can be itinerant or insulating. In the first case, the charge degrees of freedom are expected to influence the non-universal properties of the magnetic excitation spectrum. The most notorious consequence for the cuprates is the absence of outgoing magnetic branches, i.e., moving from $\tilde{Q}_\pm$ in directions that increase the distance to the commensurate wave vector $Q$. The absence of well-defined magnetic excitations in this large region of the Brillouin zone can be understood within the scenario depicted in Fig. 4. If the maximum value of $k$ for the particle-hole continuum of excitations, $k_M$, is such that $|k_M|<|\tilde{Q}_\pm|$, the magnetic excitations with $|k|>|k_M|$ and low enough energy will not be over-damped due to the absence of particle-hole excitations with an adequate energy and momentum. On the contrary, magnetic excitations with $|k|<|k_M|$ will be over-damped for any value of their energy. Another non-universal feature is the relative intensity of the peaks at the lowest-energy modes with respect to the resonance. As indicated in our work[10,11] this depends on the ratio $0 \leq \Delta_s/E_r < 1$: A small ratio hinders the detection of the resonance peak, while a large one prevents the observation of the lowest-energy branches of the magnetic spectrum.

These simple observations unify the phenomenology observed in INS of the high-$T_c$ materials. By looking at these systems we realized that these universal magnetic excitation features are inherently present in a broader class of materials we dubbed *modulated antiferromagnetic systems* (MAS[29]). This class includes conducting or insulating, long-range or disordered, commensurate or incommensurate systems. A very simple mechanism was shown to be behind its universal behaviour leading to a *twisted hour-glass-like* spectrum of magnetic excitations. Indeed, we predict that a similar twist should be seen in the insulator $La_{1.69}Sr_{0.31}NiO_4$ and in the rest of the high-$T_c$ materials whose high-energy spectrum has not been measured yet. In bilayer cuprates we expect the same qualitative behaviour for the symmetric and antisymmetric spectra of magnetic excitations since the topological characteristics of the single-particle spectrum are the same for both channels. As a result of our theory *new* predictions develop with clear challenges to experimentalists and basic constraints to a fundamental theory of high-temperature superconductivity.

**Supplementary Information** accompanies the paper on **www.nature.com/nature**.

**Acknowledgements** We thank M. Maceira for helping us with the figures, and P. Bourges for useful discussions**.**

**Competing Interests statement** The authors declare that they have no competing financial interests.

**Correspondence** and requests for materials should be addressed to C.D.B., G.O., or AVB (cdb@lanl.gov, g_ortiz@lanl.gov, avb@lanl.gov).

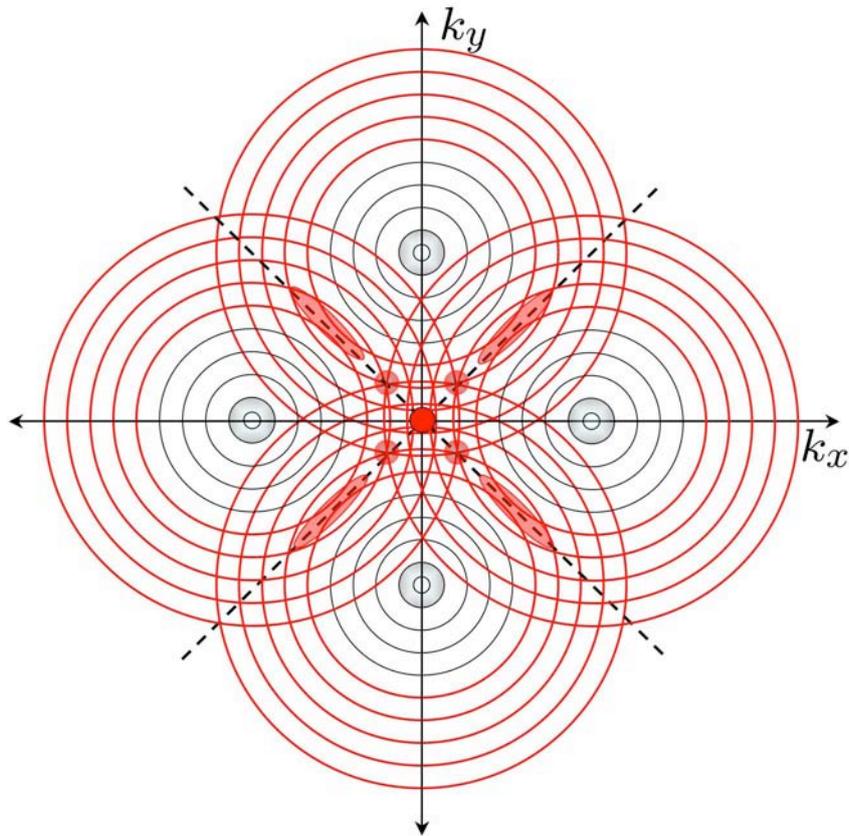

Figure 1: Illustration of the emergence of universal properties in MAS. The plane ($k_x, k_y$) represents momentum space, while the axis perpendicular to it the energy of excitations. The figure schematically shows how cones (whose increasing radius relates to increasing energy) centred at the lowest-energy (massless or massive) modes, indicated by dark circles, give rise to maxima in intensity because of a simple constructive interference as they move along the direction of increasing energy. The *resonance* peak, that lies at the centre, and the *twist* in intensity after the resonance energy (depicted as blurred circles along the diagonals) represent some of these universal properties. Therefore, these universal signatures are simply the result of a topological characteristic of the spectrum and, thus, are independent of the microscopic details of the MAS.

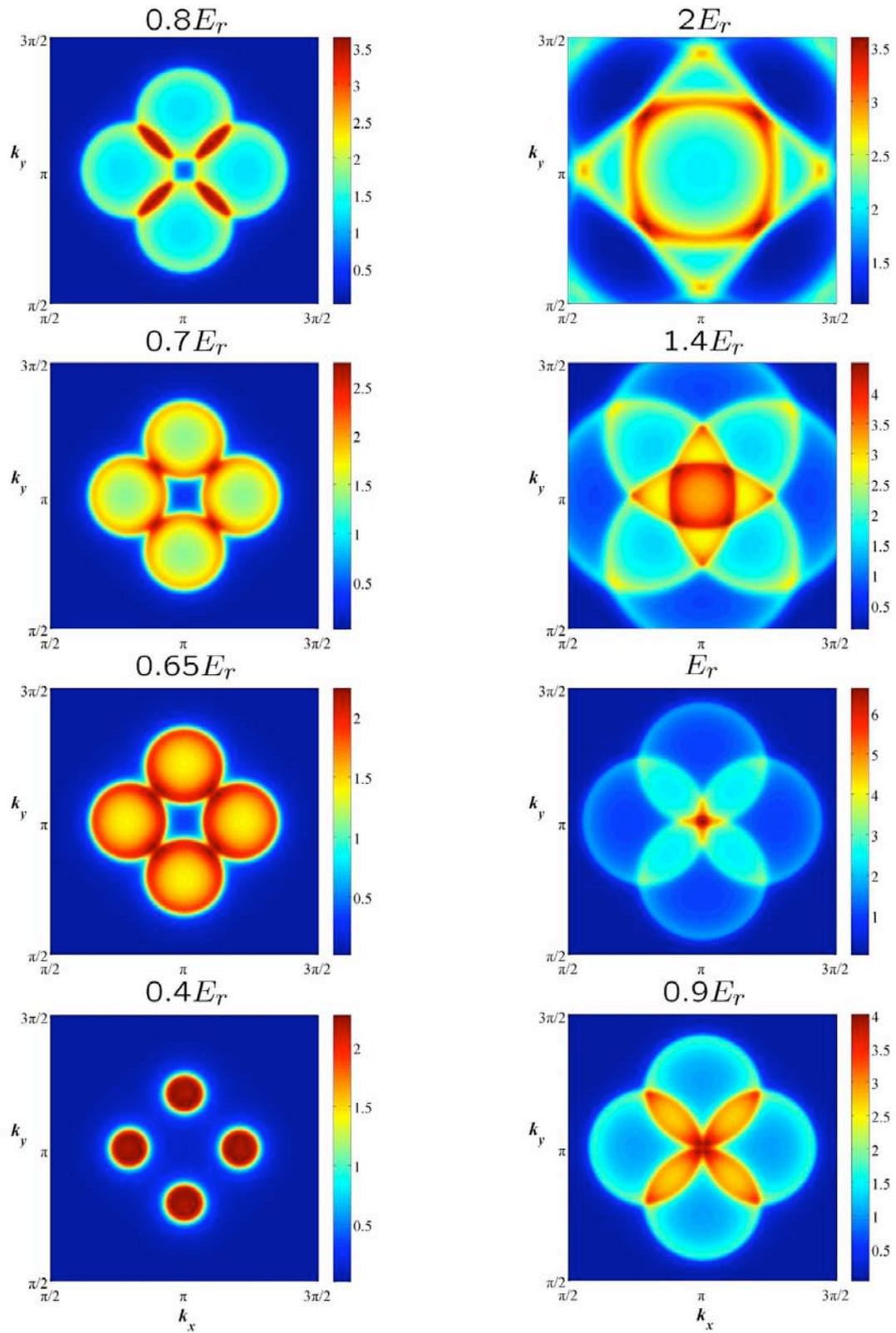

Figure 2: Constant energy contour plots of the magnetic structure factor $S(\boldsymbol{k},\omega)$

(intensity) centred around **k=Q**=($\pi,\pi$). For this particular example, we have used the Schwinger-boson mean-field result of Refs. [1,2]. Notice the way the maxima in intensity evolves as $\omega$ increases. Indeed, some of these universal features have been experimentally observed in the cuprates, while others represent our predictions. For example, there ought to be a *twist* in the location of the maxima for energies close but *smaller* than the resonance energy, which should be observable if higher experimental resolution were available.

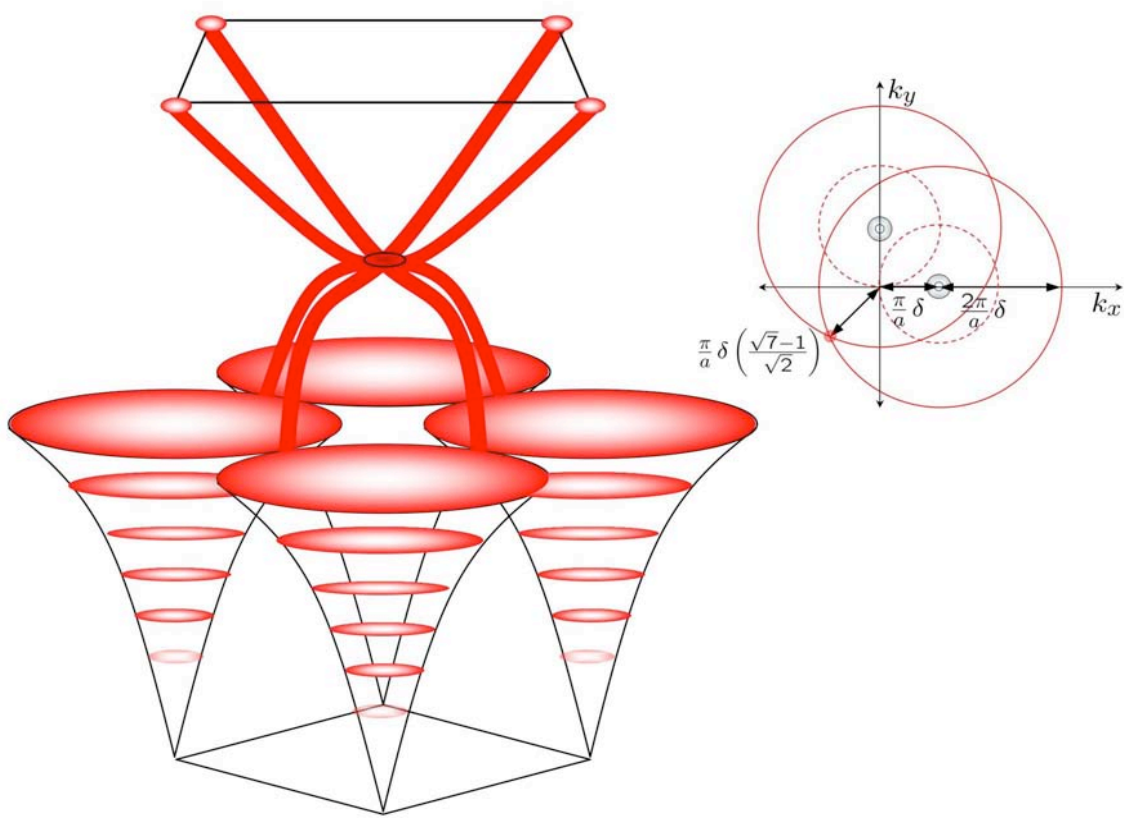

Figure 3: Three-dimensional schematics of the evolution of the maxima in the dispersion relation $\omega_{\mathbf{k}}$ (leading to a *twisted hour-glass-like* spectrum). The inset shows a simple geometric construction indicating the relation between the distance of the maxima in intensity to **k=Q**, when a linear dispersion is assumed. In particular, when $\hbar\omega = 2E_r$ the distance is $\Delta k = \frac{\pi}{a}\delta\left(\frac{\sqrt{7}-1}{\sqrt{2}}\right)$ which represents a lower bound since the real dispersion is not linear and bends downwards.

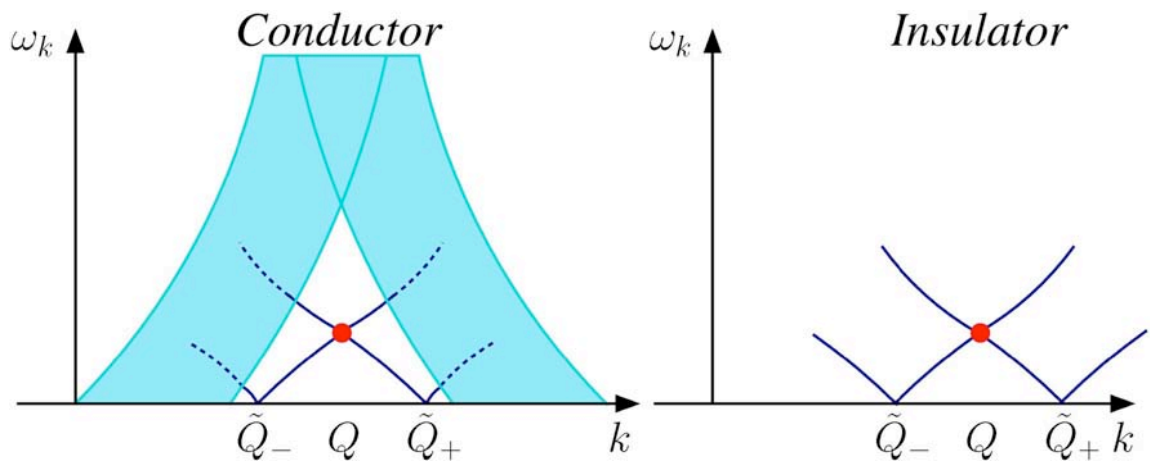

Figure 4: Example of a non-universal feature of the energy spectrum in MAS. The left panel represents a typical spectrum for a conducting material. The particle-hole continuum of excitations may *eat* the outward branches of the MAS-type spectrum, while this is not the case in an insulating material.